# ON THE ORIGIN OF LOW-SURFACE-BRIGHTNESS GALAXIES

V. ANTONUCCIO-DELOGU
Osservatorio Astrofisico di Catania
Città Universitaria - Viale A. Doria 6
95125 Catania ITALY

Theoretical Astrophysics Center
Blegdamsvej 17, DK 2100 Copenhagen, DENMARK

**Abstract** – We reconsider the problem of the origin of low-surface-brightness (LSB) galaxies within the standard Cold Dark Matter (CDM) cosmological model ($\Omega = 1, h = 0.5$). Taking into account the effect of substructure on the collapse times of low overdensity peaks, we show that the abundance of these peaks is larger than previously expected because of the dragging caused by dynamical friction induced by the presence of small-scale substructure. The number density of these shallow, low-density peaks turns out to be in good agreement with the observed overdensity of the blue LSB galaxies found during recent surveys.

## 1. INTRODUCTION

The problem of the existence of a significant amount of LSB galaxies (hereafter LSBs'), has been around since almost thirty years. Hubble (1932) and De Vaucouleur (1956) first pointed out that photographic surveys could be be biased against low-surface-brightness objects. The argument was taken up more recently by Disney (1976), who added new estimates of the relevanece of this bias on the completeness of the existing galaxy surveys. After these claims many people tried to develop new techniques to increase the sensitivity of photographic materials toward low-surface brightness levels. A new technique introduced by D. Malin allowed the serendipiditous discovery of two spectacular objects, namely Malin I and II (Bothun et al., 1987, 1990). The total HI mass of Malin 1 is about $10^{11} M_\odot$, and its disk scale-length is $\approx 55$ Kpc, making it one of the largest and shallowest spiral galaxy within the local neighborhood. Nonetheless, it could have been hardly detected at optical wavelengths due to its very low central surface brigthness (only 25.2 B-mag/arcsec$^2$, more than $4\sigma$ away from the Freeman's value of 21.65 which characterizes bright spirals). Although Malin I and II are exceptional objects under many respects, more than 20 years of efforts due to the Cardiff group (M. Disney, S. Phillips, D. Davies and collaborators) and by few other people in the USA and in Australia (most notably D. Malin, G. Impey, G. Bothun, S. McGaugh, J. Schombert) has led to the discovery of a large amount of LSBs in clusters like Hydra and in the field. These surveys have shown that, although objects as large as Malin I and II are rare, LSBs' with photometric properties more similar to those of normal galaxies but with a much smaller central surface brightness are very common. Due to the intrinsic faintness of these objects it has become possible to measure the redshifts and metallicity for some of them only recently, after the development of CCD detectors (van der Hulst et al., 1994). These first results show that these galaxies have lower-than-solar metallicities and almost no color gradients across the disc. Moreover, their HI surface density is well below the critical value for star formation in normal spirals found observationally by Kennicut (1989): this could explain the low luminosity of these objects. Although redhifts are known only for few LSBs', the intrinsic faintness of these objects strongly suggests that they cannot be detected if they were at very high redshifts. The low metallicity suggests that any interpretation based on a fading mechanism could meet serious theoretical difficulties, so we are left with the problem of explaining how galaxy formation has been delayed until recently, compared to normal spiral galaxies.

## 2. DENSITY PEAKS IN STANDARD CDM.

In this paper we will look at the predictions of a very popular cosmological model, namely the standard CDM, concerning the origin of LSBs'. This model explains quite sucessfully the observed clustering properties of the Universe from cluster up to MWBR scales but, as many other cosmological models, it runs into trouble when it tries to predict the observed photometric properties of galaxy populations. Stellar formation mechanisms, largely unknown by to-day, can hinder a significant fraction of the information contained in the initial conditions.

It has been recently suggested that environmental effects like, for instance, tidal fields, can play a significant role on star formation efficiencies, and ultimately on the observed properties of normal, bright galaxies (Lacey and Silk, 1994) The low star formation rate of LSBs' suggests that these objects could have been poorly affected by their environment, so that their photometric properties could be more directly related to the statistical properties of the density peaks from which they originated. Moreover, LSBs are found in the field and within the central regions of rich clusters like Coma, within environemnts spanning at least two order of magnitudes in density, and they seem to have the same clustering properties as bright normal spirals, although with a smaller amplitude (Mo et al., 1994). All these facts point to suggest that LSBs could be much lesser biased tracers of the post-recombination density field than normal bright spirals.

One of the main predictions of the idealized spherical collapse model (Gunn and Gott, 1972) is that only density perturbations having an initial overdensity after recombination larger than a critical value $\delta_c^{(GG)}$ ($\approx 1.68$ for a standard CDM) can turn around within lesser than one Hubble time. However, this model does not take into account the effect of the presence small-scale substructure. One of the distinctive features of the standard CDM model is a very flat power spectrum over a wide range of subgalactic mass scales ($10^6 \leq M \leq 10^9 M_\odot$). Physically this means that the Universe is very lumpy, and filled with many small-scale subclumps on these scales. In a recent series of papers (Antonuccio-Delogu, Colafrancesco and Del Popolo, 1994; Antonuccio-Delogu and Colafrancesco, 1994) we have shown that the presence of substructure induces a delay in the collapse, so that the critical overdensity for collapse is rather given by:

$$\delta_c = \delta_c^{(GG)} \left[ 1 + \frac{\lambda_0}{1 - \mu\left(\delta_C^{(GG)}\right)} \right] \quad (1)$$

Here $\lambda_0$ is a coefficient which ultimately depends on the power spectrum, and $\mu(x)$ is a known function. The dynamical friction affects more seriously the collapse of the small overdensity peaks, and it alters the statistical properties of the density fluctuation field.

This latter point can be understood by looking at the properties of the *selection function* $t(\nu)$. Following Bardeen et al. (1986) we have defined this as the probability that an average density fluctuation, selected with some prescribed filter on the scale $R_f$, will collapse on a cosmological time scale:

$$t(\nu) = \int_{\delta_c}^{\nu_{LSB}} d\rho\, p\left[\rho, \langle\bar{\delta}\rangle(R_f)\right] \quad (2)$$

where the probability distribution is given by the standard CDM (see e.g.

Table 1: Number densities of LSBs for different values of the parameters of the standard CDM. Columns are as follows: 1) filtering radius (in units of $10^2$) Kpc, 2) filtering mass (in $10^{11} M_\odot$), 3) normalization constant, 4) upper limit of central overdensity, 5) number density (Mpc$^{-3}$), 6) ratio of LSBs to total peaks' density.

| $R_f$ | $M_f$ | $b_8$ | $\nu_{LSB}$ | $n_{LSB}$ | $n_{LSB}/n_{tot}$ |
|---|---|---|---|---|---|
| 3.5 | 3.75 | 1.1 | 1   | $2.58 \times 10^{-3}$  | 0.23  |
| "   | "    | "   | 1.5 | $6.125 \times 10^{-3}$ | 0.81  |
| "   | "    | 1.8 | 1   | $8.94 \times 10^{-4}$  | 0.08  |
| "   | "    | 1.8 | 1.5 | $4.44 \times 10^{-4}$  | 0.587 |
| 2.5 | 2.73 | 1.1 | 1   | $6.71 \times 10^{-3}$  | 0.235 |
| "   | "    | "   | 1.5 | $1.6 \times 10^{-2}$   | 0.832 |
| 5.0 | 11   | 1.1 | 1   | $7.68 \times 10^{-4}$  | 0.18  |
| "   | "    | "   | 1.5 | $2.0 \times 10^{-3}$   | 0.72  |

Lilje and Lahav, 1991). It is easy to verify that the selection function has the correct asymptotic behaviour in both limits ($t(\nu) \to 1$ for $\nu \to \infty$, $t(\nu) \to 0$ for $\nu \to 0$).

If we now identify the progenitors of LSBs' with peaks of the density field having $\nu \leq \nu_{LSB}$, their number density will be given by:

$$n_{LSB} = \int_0^{\nu_{LSB}} t(\nu) N_{peak}(\nu) = \int_0^{\nu_{LSB}} d\nu \, t(\nu) \cdot \frac{1}{2\pi^2 R_*^3} \exp\left[-\frac{\nu^2}{2} G(\gamma, \gamma\nu)\right] \quad (3)$$

where the quantities in the last identity are standard within the CDM model (see e.g. Bardeen et al., 1986).

## 3. RESULTS AND CONCLUSIONS.

In Table I we show the results of applying eq. (3) for different values of $R_f$, $b_8$ (the normalization constant of standard CDM) and $\nu_{LSB}$, the maximum assumed central overdensity of those peaks giving origin to LSBs. Although the observed number density of LSBs is largely uncertain due to the incompleteness of the present surveys, recent estimates give values in the range $0.2 \leq n_{LSB}/n_{tot} \leq 0.7$ (McGaugh, 1994), in good agreement with the numbers in the last column of Table I.

We are aware of the risks of a one-to-one identification of density peaks with observed galaxy populations. In addition to stellar formation mechanisms, other features like angular momentum distribution in protogalactic clouds can induce a large variance in the observed properties. However, as we have already stressed, there are reasons to believe that environmental effects played a lesser important role on the formation of LSBs, in comparison with normal spirals. The dynamical friction induced by small-scale substructure can be seen as part of the "cross-talk" among different scales (for a review see, e.g., Silk and Wyse, 1993). At variance with other cross-talk mechanisms it is completely specified once the power spectrum is given, and does not depend on star formation. It can then be regarded as a standard ingredient of the CDM cosmological scenario.

**Acknowledgements** – I would like to express my gratitude to the referee (G. De Vaucouleur) for pointing out two references (Hubble 1932 and De Vaucouleur 1956) and for other suggestions. This work was partly supported by Danmarks Grundforskningsfond through its support for the establishment of the Theoretical Astrophysics Center (TAC).

# References

[1] Antonuccio-Delogu, V. and Colafrancesco, S., 1994, *Asrophys. J.* **427**, 72

[2] Antonuccio-Delogu, V., Colafrancesco, S. and Del Popolo, A., 1994, preprint

[3] Bardeen, J. M., Bond, J. R., Kaiser, N. and Szalay, A.S., 1986, *Astrophys. J.* **304**, 15

[4] Bothun, G.D., Impey, C.D., Malin, D. and Mould, J., 1987, *Astron. Journal* **94**, 23

[5] Bothun, Schombert, J.M., G.D., Impey, C.D. and Schneider, S.E., 1990, *Astrophys. J.* **360**, 427

[6] Disney, M.J., 1976, *Nature* **263**, 573

[7] Gunn, J.E. and Gott, J.R., 1972, *Astrophys. J.* **176**, 1

[8] Hubble, E., 1932, *Astrophys. J.* **76**, 106

[9] Kennicut, R.C., 1989, *Astrophys. J.* **344**, 685

[10] van der Hulst, J.M., Skillman, E.D., Smith, T.R., Bothun, G.D., McGaugh, S.S. and de Blok, W.J.G., 1994, *Astron. J.*, 1993, preprint

[11] Lacey, C. and Silk,

[12] McGaugh, S.S., 1994, in *Quantifying Galaxy morphology at High Redshift*, HST workshop, Baltimore MD, April 27-29 (1994)

[13] Lilje, P.B. and Lahav, O., 1991, *Astrophys. J.* **374**, 29

[14] Mo, H.J., McGaugh, S.S. and Bithun, G.D., 1993, preprint

[15] Silk, J. and Wyse, R.F.G., 1993, *Phys. Reports* **231**, 293

[16] De Vaucouleur, G., 1956, *Astron. Journal* **61**, 435